\documentclass{aa}
\usepackage{graphicx}
\usepackage{url}
\usepackage{natbib}

\def\xmm{XMM-{\it Newton}\ }
\def\sas{XMM-SAS\ }
\def\sds{SDSS~J1004+4112\ }
\def\sub{{\it Subaru\ }}
\def\epic{ EPIC\ }
\def\nh{$N_{\rm H}$\ }
\def\cxo{{\it Chandra\ }}

\begin{document}

   \title{Strange  magnification pattern in the large separation lens 
           SDSS J1004+4112 from optical to X-rays \thanks{Based on observations obtained with XMM-Newton, 
           an ESA science mission with instruments and contributions directly 
           funded by ESA Member States and NASA and observations collected 
           at the Centro Astron\'omico Hispano Alem\'an (CAHA) at Calar Alto, 
           operated jointly by the Max-Planck Institut f\"ur Astronomie and the 
           Instituto de Astrof\'isica de Andaluc\'ia (CSIC)}}

   \author{G. Lamer\inst{1}
           \and A. Schwope\inst{1}
           \and  L. Wisotzki\inst{1}
           \and  L. Christensen\inst{1,2} }

   \offprints{G. Lamer}

   \institute{Astrophysikalisches Institut Potsdam,
               An der Sternwarte 16,
               D-14482 Potsdam\\
               \email{glamer@aip.de}
               \and
               European Southern Observatory,
               Casilla 19001, Santiago, Chile}

   \date{Received; accepted}

   \abstract{We present simultaneous XMM-Newton UV and X-ray observations
             of the quadruply lensed quasar SDSS J1004+4112 (RBS 825).
             Simultaneously with the XMM-Newton observations we also performed integral 
             field spectroscopy on the
             two closest lens images A and B using the  Calar Alto PMAS spectrograph.
             In X-rays the widely spaced components C and D are clearly resolved,
             while the closer pair of images A and B is marginally resolved
             in the XMM-EPIC images.
             The integrated X-ray flux of the system has decreased by a factor of 6
             since it was observed in the ROSAT All Sky Survey in 1990, while the  
             X-ray spectrum became much harder with the power law index evolving from
             $\Gamma=-2.3$ to $\Gamma=-1.86$.
             By deblending the X-ray images of the lensed QSO we 
             find that the X-ray flux ratios between the lens images A and B
             are significantly different from the simultaneously obtained
             UV ratios and previously measured optical flux ratios. 
             Our optical spectrum of lens image A  shows an enhancement in
             the blue emission line wings, which has been observed in previous epochs as 
             a transient feature. We propose a scenario where intrinsic UV
             and X-ray variability gives rise to line variations which are 
             selectively magnified in image A by microlensing. 
             The extended emission of the lensing cluster of galaxies is clearly detected 
             in the \epic\ images, we measure a 0.5-2.0 keV  luminosity of $1.4 \times 10^{44}{\rm erg/s}$.
             Based on the cluster X-ray properties, we estimate a mass of $2-6 \times 10^{14} {\rm M_\odot}$.

            }

     \authorrunning{G. Lamer et al.}
     \titlerunning{Strange magnification pattern in SDSS J1004+4112}

     \maketitle

\section{Introduction}

The lensed QSO SDSS J1004+4112 was originally  found as an X-ray source
in the ROSAT All Sky Survey (RASS) and  entered the ROSAT Bright Source catalogue
under the designation RBS 825 \citep{schwope}. The brightest optical counterpart 
within the RASS error circle turned out to be a type I QSO at z=1.73
and was regarded as the unique identification of the unresolved X-ray source.

Its lensed nature was discovered in a survey of large separation lenses 
using the Sloan Digital Sky Survey (SDSS) \citep{inada*:03}. Optical imaging 
revealed 4 images of the quasar, with a maximum separation of $14.6''$ it is
the largest separation  lensed QSO known so far. A faint 5th image  was  
detected in deep HST imaging by \citet{inada05}. 
The lensing object is
a cluster of galaxies at z=0.68, the quasar itself has a redshift of z=1.73.
SDSS J1004+4112 is the first multiple quasar where the lensing gravitational potential
is dominated by a cluster of galaxies, and not a single galaxy.
Modelling of the lensing potential gives only a rough estimate of the cluster mass
($M\geq 10^{14} h^{-1} M_{\odot}$, Oguri et al. 2004).
Similarly the predictions for the time delays between the images are uncertain,
the delay between the closest components A and B can be up to $37\, h^{-1}$ days, the largest 
delay C-D could be up to  $3000\, h^{-1}$ days. Most models of the lens predict 
a  delay sequence C--B--A--D of the images, but the  ordering D--A--B--C is also
possible \citep{oguri}.
Optical spectroscopy of the individual components revealed significant differences in the
emission line profiles of the components: the blue wings 
of the emission lines were found to be  enhanced in the spectra of component A \citep{richards*:04:ML} 
at certain epochs in 2003. This was interpreted as
an indication for the presence of microlensing in lens image A.
In this paper we present simultaneous multi-waveband observations of \sds from optical 
spectroscopy to UV and X-ray  observations with \xmm.
In Section \ref{sect:xmmobs} we describe the analysis of the \xmm data, the results 
are compared with an earlier epoch ROSAT observation in Section \ref{sect:rosat}. 
In Section \ref{sect:discussion} we attempt an explanation of the unusual spectral 
energy distributions of the lens images (see Section \ref{sect:sed}) 
and the equally puzzling differences 
in the emission line profiles between the QSO images A and B (see Section \ref{sect:pmas}).

If not stated otherwise, all statistical error values given throughout the paper are $1 \sigma$ limits. 
All luminosities were calculated using the cosmological parameters $H_0=70\; {\rm km\; s^{-1} Mpc^{-1}}$,
$\Omega_{\rm M}=0.27$, and $\Omega_{\Lambda}=0.73$

\section{XMM Newton observations}
\label{sect:xmmobs}

\object{\sds} was observed by \xmm on 20/04/2004 under observation ID 0207130201 for 
a total of 60 ksec. 
The primary instruments were the 3 \epic (European Photon Imaging Camera) imaging CCDs, observing
simultaneously with the optical monitor (OM) and the RGS grating
spectrometers. A summary of the \epic and OM observations is given in 
Table \ref{obssum}. The RGS spectra do not show sufficient signal for 
scientific analysis and are not discussed here.

\begin{table}[htb]
\caption{\label{obssum}  Summary of \xmm observations}
\begin{tabular}{llccr}\\ \hline
Instr. & Filter & Date        &  Start UT & exp. [s] \\
 \hline
EMOS1      & Thin1  & 2004-04-20  & 02:59:04       &  60473 \\
EMOS2      & Thin1  & 2004-04-20  & 02:59:03       &  60478 \\
EPN        & Thin1  & 2004-04-20  & 03:21:24       &  58836 \\
OM         & U      & 2004-04-20  & 03:03:00       &  8000  \\
OM         & UVW1   & 2004-04-20  & 04:58:08       &  9370  \\
OM         & UVM2   & 2004-04-20  & 04:58:08       &  15364 \\ 
\hline
\end{tabular}
\end{table}

\subsection{X-ray data analysis}

The data were processed from observation data files (ODFs) using
the XMM Science Analysis Software (SAS) version 6.5. Part of the \epic exposure time was affected by high
particle background.
Since the background contamination is not severe and \object{\sds} is a relatively bright
X-ray source, we decided not to screen out the times of enhanced background but to use all available data.
The images were binned with 2 arcseconds per pixel in order to optimally sample the 
5 arcsec (FWHM) PSF of the XMM telescopes.
For each of the 3 \epic cameras we created 5 images in the energy bands listed in table \ref{bands}.

\begin{table}[h]
\caption{\label{bands}  \epic image energy bands}
\begin{tabular}{ll}\\ \hline
band 1 & 0.2-0.5 keV \\
band 2 & 0.5-1.0 keV \\
band 3 & 1.0-2.0 keV \\
band 4 & 2.0-4.5 keV \\
band 5 & 4.5-12.0 keV \\
\hline
\end{tabular}
\end{table}

The corresponding exposure maps for each image and detection masks
were calculated using the {\sas} tasks {\tt eexpmap} and {\tt emask}. 
The background maps were constructed using the SAS task {\tt esplinemap} 
by fitting a spatially flat component plus a component following the vignetting function
of the X-rays telescopes to source-free regions of the EPIC images.

\begin{figure}[htb]
  \includegraphics[width=8.5cm]{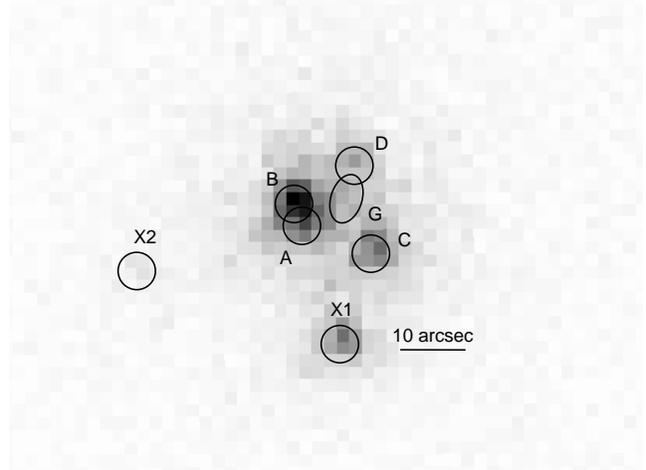}
  \caption{Combined \epic MOS image (0.2-4.5 keV) of \object{\sds}
           with positions of sources used in PSF fitting.}
  \label{mosimage}
\end{figure}

\begin{figure}[htb]
  \includegraphics[width=8.5cm]{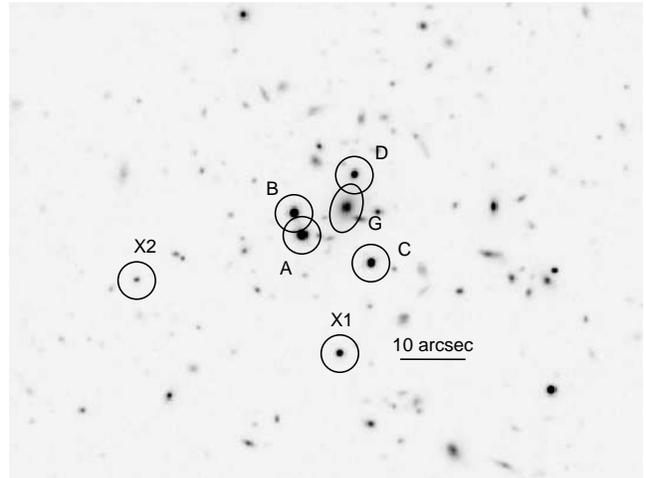}
  \caption{Archival subaru I-band image of \object{\sds} with source positions as in Fig. \ref{mosimage}.}
   \label{subaru}
\end{figure}

\subsubsection{PSF fitting and extended emission from the lensing cluster}

To  measure the X-ray flux of each of the partly confused
lens components and to search for the presence of diffuse X-ray emission
from the lensing cluster, we performed multi-PSF fitting on the XMM \epic X-ray images. 

The physical pixel size of the 2 \epic MOS cameras is 40 $\times$ 40 ${\rm \mu m}$,
corresponding to 1.1 arcsec. 
The \epic PN pixel size of 150 $\times$ 150 ${\rm \mu m}$ corresponds to 4.1 arcseconds,
undersampling the PSF of the X-ray telescope. 
Since the angular distance between the closest components of \object{\sds} is 
3.8 arcsec, slightly smaller than the FWHM of the XMM XRTs, only the well sampled 
images of the \epic MOS cameras are used for the PSF fitting described here.

We have used the XMM SAS task {\tt emldetect} to perform multi-PSF fitting on 
\epic MOS images. The version of {\tt emldetect} from SAS release 6.5 was modified 
to use the ``extended accuracy'' calibration PSF.
The ``medium accuracy'' PSF model, which is normally used by {\tt emldetect} is the only 
calibration PSF model, which includes the azimuthal asymmetries of the XMM PSF at large off-axis
angles. However, near the optical axis, the ``extended accuracy'' PSF is a more accurate
representation of the radial brightness profile. 
Any spatial variations of the exposure maps and the background maps are incorporated 
in the fitting model.

Apart from the four lensed components of \object{\sds}, another two X-ray point sources are 
visible in the immediate vicinity of \object{\sds}. These sources, labelled with {\it X1} and {\it X2}
in Fig. \ref{mosimage}, can also be identified with optical counterparts in \sub images
(see Fig. \ref{subaru}) and are presumably unrelated AGN. Object {\em X1} has been used to tie 
the astrometry of the XMM 
image to the astrometric frame of the \sub images. 

The optical positions of the 4 lens components, and the objects {\it X1} and {\it X2},
were then used as fixed input positions for multi-PSF fitting of point-like sources. 
The fits were performed on a co-added MOS1+MOS2 broad band image in the energy range 0.2-4.5 keV 
as well as on a set of 5 images in the bands described in Table \ref{bands}.

After the subtraction of the 6 point sources with their best fit fluxes
it became obvious that extended emission from the lensing cluster of galaxies 
is present in the X-ray images.
Therefore we added an extended component (the instrumental PSF convolved with a
 King profile) to the fitting procedure. The core radius and 
the position of the extended source was left free to vary. This model was fitted to the 
broad band EPIC MOS image.

The best fit position of the extended source is $\alpha$=10h04m33.9, $\delta$=41d12m50.6s, 
close to quasar image D. The resulting core radius is $r_c = 30''$. 
We used these values as fixed input 
parameters to obtain fluxes for the cluster and the point sources in each of the 5 
X-ray energy bands. The results are summarized in table \ref{rates}.
The  point source fluxes resulting from these fits are also used in the discussion of the 
X-ray properties of the lens images and are the basis of the spectral energy
distributions shown in Figs. \ref{f:sednfn} \& \ref{f:sedrel}

\begin{figure}[htb]
  \includegraphics[width=8.5cm]{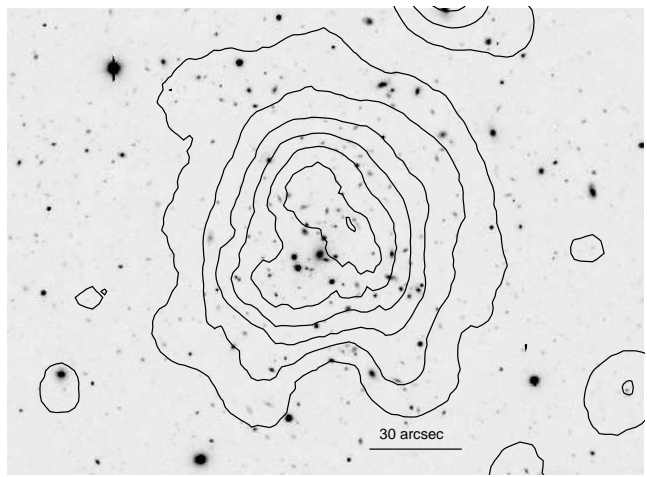}
  \caption{Subaru I band image with  X-ray contours from the combined MOS images.
           The point sources were subtracted with fluxes as listed in table \ref{rates}. }
   \label{residual_contours}
\end{figure}

We measure a 0.5-2.0 keV luminosity of $1.4 \times 10^{44} {\rm erg\;s^{-1}}$.
Fig.\ref{residual_contours} shows that the extended cluster emission is asymmetric with respect to the
brightest cluster galaxy and the lensing centre.

From the count rates in the 5 standard \epic MOS energy bands we constructed
a 5-point spectrum which we fitted with a 
{\tt mekal} model 
at the cluster redshift of z=0.68 and a fixed metallicity of $Z=0.5$, which was measured for 
 clusters of this redshift range \citep{tozzi}.
We derived a best fit gas temperature of $T=4.3^{+2.1}_{-1.0} {\rm keV}$.
With the scaling relations for clusters at  this redshift (Kotov \& Vikhlinin, 2005)
its luminosity and temperature puts the cluster in the mass range $2-6 \times 10^{14}  {\rm M_{\odot}}$,
consistent with, and improving the estimates from lensing models.

The best fit centroid of the cluster extended emission is located 7 arcseconds 
northwest of the brightest cluster galaxy. The offset well  exceeds
the statistical $1 \sigma$ positional error (0.4 arcsec) of the extended X-ray source.  
However, the \cxo image of \sds \citep{ota} shows a peak of the 
diffuse X-ray emission at the position of the BCG. 
The discrepancy can be understood, since due to dominance of the QSO point sources 
at the centre of the cluster, the 
fit to the EPIC MOS image will be dominated by the outer regions of the cluster emission.
Indeed, Fig. \ref{subaru} shows, that the outer contours of the X-ray emission are offset from the
BCG to the northwest.  This may be an indication that the cluster is not in a fully relaxed state. 

\begin{table*}[htb]
\caption{\label{rates} Fluxes}
\begin{tabular}{cccccc}\\ \hline
band       &  \multicolumn{5}{c}{flux of component}\\   
           &   A             &    B             &    C             &   D              &   cluster      \\
 keV     &              \multicolumn{5}{c}{\rm $10^{-15} {\rm erg /(cm^2 s)}$}                                                \\
 \hline    
 0.2-0.5   & $ 6.7\pm1.2$  & $21.7\pm1.3$  &  $11.8\pm1.0$    &  $4.8\pm0.8$  & $17.8\pm3.3 $  \\ 
 0.5-1.0   & $19.3\pm1.8$  & $29.5\pm1.8$  &  $20.9\pm1.4$    & $12.2\pm1.2$  & $31.1\pm4.8 $  \\
 1.0-2.0   & $21.1\pm2.0$  & $38.1\pm2.0$  &  $25.6\pm1.6$    & $13.7\pm1.4$  & $62.1\pm5.9 $  \\
 2.0-4.5   & $17.9\pm3.7$  & $63.6\pm3.9$  &  $40.0\pm3.2$    & $24.3\pm2.8$  & $46.8\pm12.0$  \\
4.5-12.0   & $31.2\pm6.3$  & $33.1\pm6.2$  &  $60.5\pm6.4$    & $21.9\pm4.6$  & $22.7\pm24.8$  \\
\hline
\end{tabular}
\end{table*}

\subsubsection{XMM Optical monitor data}

The XMM Optical Monitor (OM) is a 30 cm telescope equipped with photon counting
micro-channel plate intensified CCDs (MICs). 
In imaging mode various filters covering passbands from  180 nm to 600 nm can be 
used. The FWHM of the OM PSF ranges from 1.4 arcsec to 2.0 arcsec, depending on the
filter. 
Due to the nature of the detectors, the measured event rates have to be corrected
for the effects of dead time, coincidence losses (in case of bright sources),
and a slow degradation of the detector sensitivity over the time of the mission.

\begin{figure}[htb]
  \includegraphics[width=8.5cm]{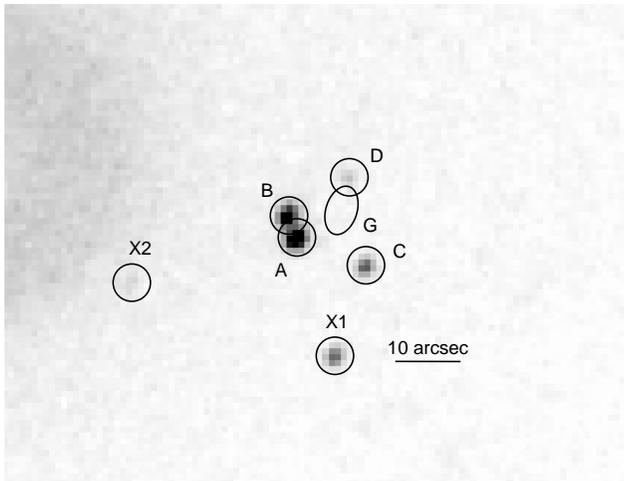}
  \caption{XMM OM U-band image of \object{\sds} with source positions as in Fig. \ref{mosimage}.}
   \label{om_u}
\end{figure}

We have images of \object{\sds} using the OM filters U, UVW1, UVM2 with central wavelengths
344 nm, 291 nm, and 231 nm (see Fig. \ref{om_u} for the OM U-band image).
The data were processed with the OM reduction pipeline {\tt omichain}
from SAS version 6.5.
Since in the OM images the PSFs of the lens components  overlap,
we used PSF fitting to measure the fluxes of the components.
For this purpose the backgrounds of the OM images, including the OM-typical
 reflection features were modelled by fitting low-order bivariate polynominal
surfaces.
On the background subtracted images we fitted the lens components and the 
close-by source 'X1' using the  {\tt GALFIT} package \citep{peng}. 
None of the cluster galaxies was detected in the OM images.  
We used a Gaussian-shaped PSF with freely variable width, 
forced to an identical value for all sources.
This method does not necessarily produce exact values for the absolute
fluxes, if the true PSF deviates significantly from the Gaussian model.
However, the relative fluxes of the components, which are more important
in this context, will be  measured with high accuracy.

The resulting count rates were corrected for detector dead time
and for the time-dependent degradation factor, as computed by the  
SAS task {\tt ommag}. For sources in the brightness range of \object{\sds} the count 
rate dependent coincidence loss correction is of the order of 1\% and 
has been neglected here. Table \ref{om_tab} shows a summary of the
OM results.

\begin{table*}[htb]
\caption{\label{om_tab} OM UV data}
\begin{tabular}{lllllll}\\ \hline
QSO image &  \multicolumn{3}{c}{raw rate}  &  \multicolumn{3}{c}{flux density}    \\
          &  \multicolumn{3}{c}{[cts/sec]} &  \multicolumn{3}{c}{$\rm{10^{-16} erg/(cm^2 s \,\AA)}$}    \\
          &    U &  UVW1  & UVM2           &  (344 nm)  &  (291 nm)   &  (231 nm) \\ \hline
A         & 1.109 & 0.441 & 0.078       & 2.24 & 2.24 &  1.87  \\
B         & 0.782 & 0.305 & 0.050       & 1.58 & 1.55 &  1.20  \\ 
C         & 0.348 & 0.117 & 0.018       & 0.702 & 0.60 &  0.43 \\
D         & 0.111 & 0.048 & 0.005       & 0.224 & 0.25 &  0.13 \\
\hline
\end{tabular}
\end{table*}

\subsubsection{X-ray spectra}

The lens images  A and B are too close to extract individual \epic spectra of these components.
Therefore for each \epic camera we extracted individual spectra for the components C and D
as well as for the source blends A+B and for the entire complex of  the 4 QSO images. 
We used source extraction radii of 4 arcseconds for the single components and 8 arcseconds
for the A+B blend.
The extraction region for the entire complex with 16 arcseconds radius encloses all 4 lens images and the core
of the lensing cluster, but not  
the unrelated X-ray source 'X1'.

The background was extracted from
source-free regions. 
Detector response matrices and effective areas were computed using the software and calibration files
from \sas version 6.5. 
Spectral fits have been performed with {\tt XSPEC} v22.1. For all spectral fits we used a single power law model with
photoelectric absorption and  an optional  Fe K$_{\alpha}$ fluorescence line  at 6.4 keV (AGN source frame).

\begin{figure}[htb]
  \includegraphics[angle=270,width=8.5cm]{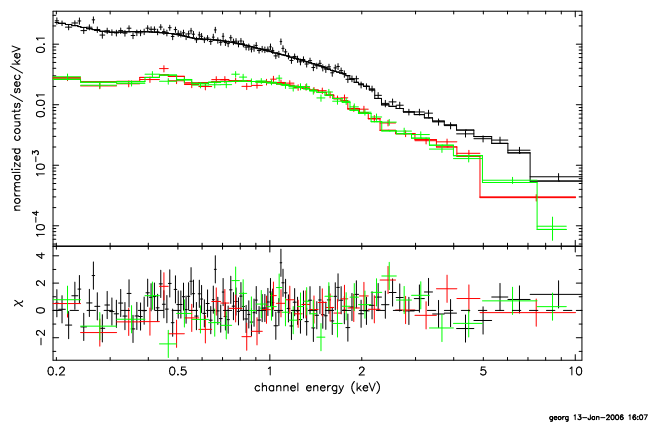}
  \caption{\epic XMM MOS and PN spectra fitted with a single power law model. At
           the rest frame energy 2.3 keV a weak signature of the Fe K$_{\alpha}$
           line is visible. The upper spectrum (black) is the PN spectrum with the measured 
           data points and a line indicating the best fit model. 
           The lower spectra (red and green) show the MOS1 and MOS2 data with the same model. 
           See the online edition of the journal for a colour
           version of this figure.}
   \label{spectrum}
\end{figure}

The global spectrum including all lens images is fitted well by the single power law model with
photon index $\Gamma=-1.86 \pm 0.02$ and photon flux density $F_{\rm 1 keV} = (6.50 \pm 0.18) \times
10^{-5}$\, cm$^{-2}$\,s$^{-1}$\,keV$^{-1}$ . 
With a best fit absorbing column density $N_{\rm H}= (1.04\pm 0.19) \times 10^{20} {\rm cm}^{-2}$
there is no indication for  intrinsic or other line-of-sight absorption 
in excess of the galactic column density $N_{\rm H,gal}= 1.14 \times 10^{20} {\rm cm}^{-2}$ \citep{dickey}. 
Fig. \ref{spectrum} shows the EPIC spectra with the best fit power law model.
The inclusion of a 6.4 keV iron fluorescence line at the redshift of the QSO does not improve the
$\chi^2$ significantly, the best fit results in an unresolved line with  equivalent 
width $55\pm 39$ eV at the source rest frame energy $6.48 \pm 0.06$ keV.

The results of the spectral fits for the individual lens images are summarized in Table \ref{spectab}.
None of the spectra show evidence for excess absorption. 
When we include a narrow Fe emission line at the source frame energy 6.4 keV, the detection 
is marginal for images A+B,  only upper limits can be given for the line equivalent width in images C and D.

\begin{table}[htb]
\caption{\label{spectab}  XMM spectra of individual components: }
\begin{tabular}{lcccc}\\ \hline
       &           \multicolumn{2}{c}{free \nh}    & gal.\nh   &  Fe$ K_{\alpha}$  \\ 
Image  & \nh                          &    $\Gamma$    &     $\Gamma$   &  EQW       \\
       &     $[10^{20}{\rm cm^{-2}}]$ &                &                &      [eV]  \\
\hline
A+B    &  $0.55 \pm .26$         & $1.87 \pm .03$  &  $1.92 \pm .02$ &  $ 97 \pm 59$       \\
 C     &  $0.53 \pm .54$         & $1.78 \pm .05$  &  $1.83 \pm .03$ &  $ < 82 (90\%) $  \\
 D     &  $0.17 \pm .84$         & $1.69 \pm .07$  &  $1.75 \pm .04$ &  $ < 162 (90\%)  $ \\
\hline
\end{tabular}
\end{table}

\subsubsection{Chandra spectra}

During the refereeing process of this article,
results of a \cxo observation of \sds were
published as a yet unrefereed preprint \citep{ota}.
The \cxo observation was performed in January 2005, some 9 month
after our \xmm observation. 
We give a brief comparison of the \cxo and \xmm results 
in Sect. \ref{sect:discussion}.
However, Ota et al. do not give the  power law spectral indices
of the individual QSO images.
Therefore we extracted spectra of each QSO component from the 
archival \cxo data using CIAO version 3.3 and CALDB version 3.2.1.
For each component we applied a source extraction radius of 1.75 arcseconds 
and chose a source free backround region.
These spectra were fitted with single power law spectra with fixed galactic absorption.
The resulting spectral indices (Table \ref{chandra}) can directly be compared with
the corresponding values measured by \xmm (Table \ref{spectab}).
For all QSO images, the spectral indices measured by \xmm and \cxo are consistent 
with each other. During both observations, image D exhibited a harder spectrum than the other
components.

\begin{table}[htb]
\caption{\label{chandra}  Chandra spectral indices of individual components: }
\begin{tabular}{lc}\\ \hline

Image  & $\Gamma$  \\

\hline
A    &  $ -1.87 \pm 0.06 $  \\   
B    &  $ -1.93 \pm 0.05 $  \\
C    &  $ -1.92 \pm 0.07 $  \\     
D    &  $ -1.67 \pm 0.07 $  \\   
\hline
\end{tabular}
\end{table}

For all QSO images \citet{ota} give Fe $K_{\alpha}$ equivalent widths that are much
larger than the corresponding values or  upper limits from the (higher SNR) \epic spectra (Table \ref{spectab}).
Since there is considerable uncertainty about the \cxo effective area calibration at energies around 
2.3 keV\footnote{see \url{http://cxc.havard.edu/ciao3.3/why/caldb3.2.1_hrma.html}}, where the redshifted line is measured,
we do not regard this discrepancy as evidence for any spectral variability.

\subsubsection{X-ray light curves}

For all \epic cameras we also extracted the  events detected from the complex
of images A and B with an extraction
radius of 6.5 arcseconds and created background subtracted 
light curves with 1000 seconds bin size. 
The light curves do not show any indication for variability within the
16 hours interval of the observation.

\section{ROSAT X-ray observations}
\label{sect:rosat}

The region of \object{\sds} was scanned between Oct 17 and Nov 9, 1990, with the ROSAT
X-ray satellite during its all-sky survey (RASS). 
It received a total exposure of 474
seconds; 112 source photons were detected from the region of \object{\sds}. The maximum
likelihood fit applied to the RASS-data was slightly indicative of a small
extent (Gaussian $\sigma$ of 9\arcsec\ in excess of the RASS-PSF of about
20\arcsec) of the source, however with a very low extent likelihood. There are
no further observations of this region of the sky with ROSAT.

The ROSAT X-ray hardness ratios HR1$ = -0.32\pm 0.09$ and HR2$=0.15\pm0.16$
lie well in the range which are typically populated by AGNs of different
flavours. The hardness ratios are defined as HR$=\frac{H-S}{H+S}$, with $H$
and $S$ being the counts in soft and hard energy channels, respectively. For
HR1 and HR2 the relevant soft channels are $0.1 - 0.4$ and $0.5 - 0.9$\,keV,
respectively, and the relevant hard channels are $0.5 - 2.0$\,keV and $0.9 -
2.0$\,keV. 
For a more detailed spectral analysis we extracted source photons from the
X-ray event lists using the MIDAS/EXSAS software. A background-corrected
source spectrum was binned into 5 independent energy bins with a minimum
signal-to-noise ratio of 5 per bin. The spectrum could be successfully fitted
with a simple power law absorbed by  cold interstellar matter with the column
density fixed at the galactic value, $N_{\rm H} = 1.14 \times
10^{20}$\,cm$^{-2}$. The best fit with reduced $\chi_\nu^2 = 0.2$ for 3 degrees of
freedom was achieved with a power-law index $\Gamma = -2.3\pm0.3$ and a
photon flux density at $E_0 = 1$\,keV of $F_{\rm E_0} = (4.0 \pm 0.8) \times
10^{-4}$\, cm$^{-2}$\,s$^{-1}$\,keV$^{-1}$.

Compared to the spectrum taken by \xmm\ 14 years later, the RASS X-ray spectrum
of \object{\sds} is much brighter and softer (see Fig. \ref{f:sednfn}). 
X-ray variability between bright-soft and low-hard states has been observed 
in other type I AGN with a similar range of spectral slopes (e.g. \object{NGC~4051}, Lamer et al. 2003).

\section{Spectral energy distribution of the lens images}
\label{sect:sed}

\begin{figure}
\resizebox{\hsize}{!}{\includegraphics[angle=-90]{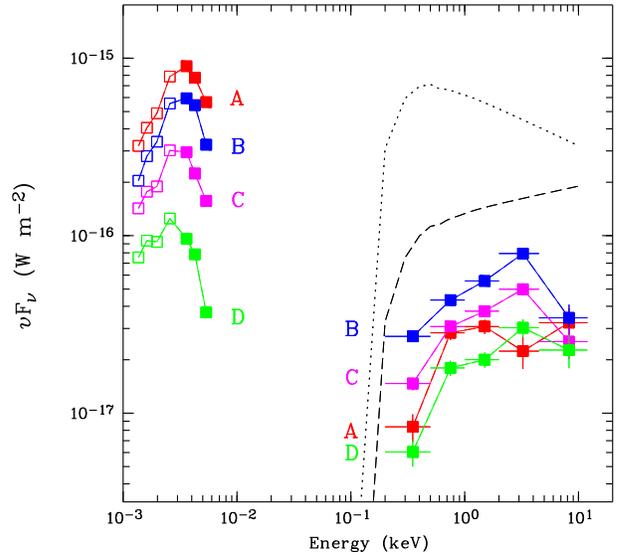}}
\caption{Spectral energy distributions (SEDs) 
of the lensed components of \object{\sds} from the
optical to the X-ray wavelength regime in $\nu F_{\nu}$. 
The simultaneous \xmm measurements are marked with filled squares, 
the earlier epoch Subaru data points are indicated by open diamonds.
The lines represent the best-fit power laws for the integrated light of
the lens and the QSO based on ROSAT all-sky survey (dotted) and XMM-Newton (dashed)
observations. See the online edition of the journal for a colour
           version of this figure.}
\label{f:sednfn}
\end{figure}

\begin{figure}
\resizebox{\hsize}{!}{\includegraphics[angle=0]{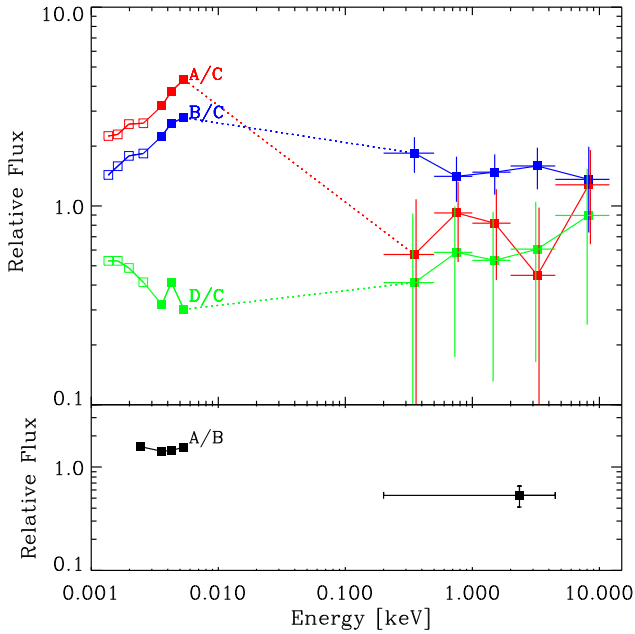}}
\caption{Top panel:  SEDs of QSO components A,B, and D from the optical 
         to the X-ray regime as in Fig. \ref{f:sednfn}, but relative to component C.
         Bottom panel: SED of component A relative to component B. Here only
         simultaneous data are used: the continuum flux ratio from the optical spectrum,
         the XMM OM UV data, and the XMM EPIC X-ray data. See the online edition of the journal for a colour
           version of this figure.}
\label{f:sedrel}
\end{figure}

With the results of the PSF fitting to the \epic MOS images together
with the results from the simultaneous XMM OM imaging and the earlier-epoch Subaru
photometry \citep{oguri} we constructed spectral energy distributions (SEDs) for each lens image.
Figure \ref{f:sednfn} shows a comparison of the SEDs together with integral spectra of the 
whole lens complex from the XMM observation and the ROSAT all-sky survey observation.
Figure \ref{f:sedrel} shows relative SEDs of lens images A,B,D normalized to the SED of image C.

It is obvious that the SEDs  deviate significantly from the simple assumption of achromatic lensing
of a constant source.
Image A shows a strong deficit in X-ray flux, while images A and B differ strongly
from images D and C in the flux and slope of the UV spectrum.

Dust extinction in intervening absorbers could alter the UV flux
ratios of the QSO images. We therefore estimate its possible effect
on image D, which has the reddest optical to UV  SED and whose optical
spectrums shows strong \ion{Mg}{ii} absorption lines near z=0.7.
When we include a photoelectric absorber at z=0.7 in addition to 
fixed galactic absorption when fitting the \epic spectrum of image D,
we derive a $90\%$ upper limit of $2 \times 10^{20} {\rm cm^{-2}}$ for
its column density.
According to  the galactic $E(B-V)/N_{\rm H}$ relation by \citet{bohlin}, we
estimate an upper limit of $E(B-V)=0.034$ in the frame of the absorber.
Using the exintinction laws from \citet{cardelli}, we arrive at an upper limit
for the extinction $A(200 {\rm nm})=0.3$. Redshifted by z=0.7, this wavelength
corresponds to the OM U-band, where an extinction by 0.6 mag would be  required
to explain the D/C ratio.
We therefore conclude that extinction on the line of sight is not able to 
account for all the  wavelength dependence in the D/C flux ratio.
Since the SEDs of images A and B differ even stronger from that of
image D, we regard intrinsic variability of the QSO  as the most likely
cause for the deviations in optical/UV SEDs.

The comparison of the RASS spectrum with the global XMM spectrum shows that \object{\sds}
was in a much brighter and softer X-ray state during the epoch of the ROSAT observations.
We therefore can assume a strong intrinsic variability of the quasars X-ray emission.

Comparing the X-ray flux ratio A/B ($0.53\pm0.12$, Fig. \ref{f:sedrel}) with the value measured by \cxo 
9 month later (${\rm A/B}=0.79\pm0.03$, Ota et al., 2006), we find a marginally significant 
variation at the 2 $\sigma$ level.   
 
\section{Optical spectroscopy}
\label{sect:pmas}

\begin{figure}
\includegraphics*[angle=-90,width=\linewidth]{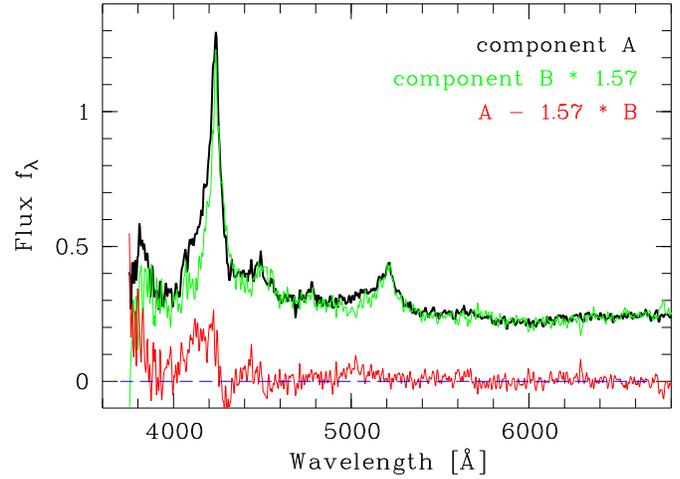}
\caption[]{Optical spectra of \object{\sds} A and B, obtained with PMAS 
on Calar Alto. For the purpose of comparison, the fainter component B 
has been scaled to match component A in the continuum. Notice the
substantial mismatch in the blue wings of the emission line profiles.
See the online edition of the journal for a colour version of this figure.}
\label{fig:specAB}
\end{figure}

\begin{figure}
\includegraphics*[angle=-90,width=\linewidth]{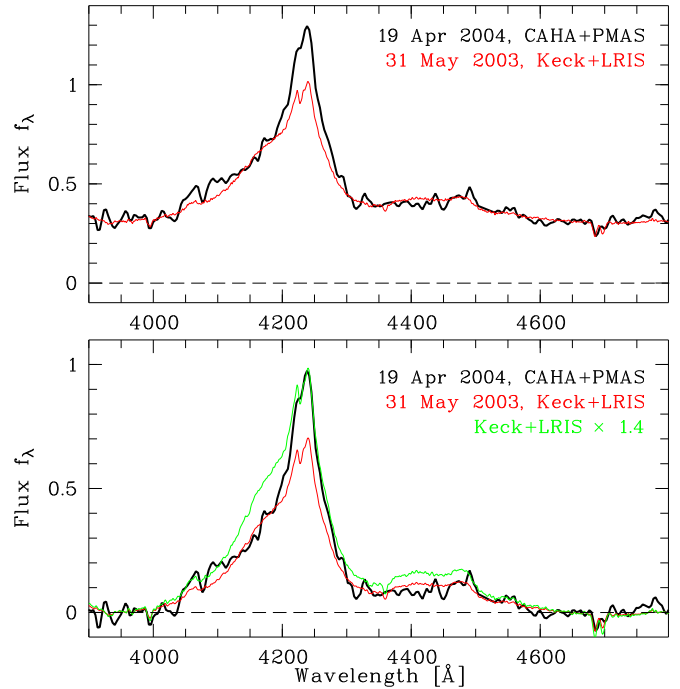}
\caption[]{Top panel: Comparison of spectra in the region around the \ion{C}{iv} 
emission line of component A between May 31, 2003 (spectrum kindly provided by 
Dr.\ G.T.~Richards) and Apr 19, 2004. 
The almost perfect match in the continuum between the two epochs is 
not by design and possibly accidental.
Bottom panel: Same data, but after subtraction of a local
pseudo-continuum, thus underlining the comparison of emission line
\emph{profiles}. Two different scaling factors were explored (see text).
See the online edition of the journal for a colour
version of this figure.} 
\label{fig:specA_PK}
\end{figure}


\begin{figure}
\includegraphics*[angle=-90,width=\linewidth]{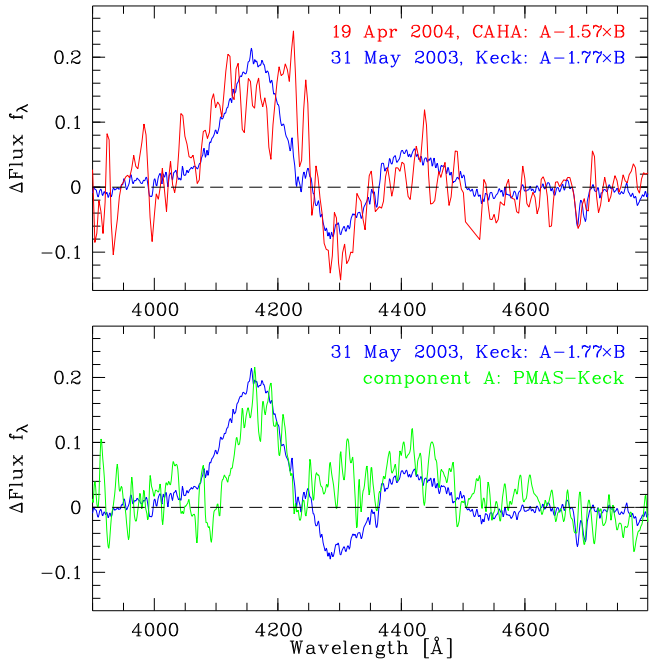}
\caption[]{Top panel: Comparison between $f(A) - S\times f(B)$
emission line residuals in the \ion{C}{iv} spectral region. Red line: 
Difference between \object{\sds} A and B in the PMAS spectra. Blue line: 
The same for the 2003 Keck spectra. 
Bottom panel: Green line shows
the difference between PMAS and Keck data of component A, compared to
the $f(A) - S\times f(B)$ residual for the 2003 Keck spectrum.
Please see the online edition of the journal for a colour version
of this figure.}
\label{fig:specA_resid}
\end{figure}

\subsection{Observations and data reduction}

We targeted \object{\sds} during an observing run on 19 April, 2004, using 
the \emph{Potsdam Multi-Aperture Spectrophotometer} PMAS,
mounted at the Calar-Alto 3.5~m telescope
\citep{roth*:05:PMAS}.
PMAS features a $16\times 16$ elements microlens array 
coupled to the spectrograph by optical fibres. 
The image scale is $0\farcs 5$ per spatial pixel (\emph{`spaxel'}).
Because of the limited field of view of $8''\times 8''$
we only observed components A and B (which fitted easily into 
one pointing).
The spectral resolution was $\sim 6$\,\AA\ FWHM, the spectra cover the
wavelength range 3800--6800 \AA. 

Four subsequent exposures of 30~min were obtained. Each was reduced
with the procedure described below, after which the spectra were coadded
with inverse variance weighting. As the atmospheric transmission during
that night turned out to be rather variable, the data are not of 
absolute spectrophotometric quality. Before coaddition, all exposures were
therefore rescaled to match the spectrum with the highest count level.
We applied a formal spectrophotometric calibration 
obtained from observing the standard star HR~3454 immediately 
before the QSO. Notice that the \emph{relative spectrophotometry} 
between components A and B is still very accurate, as there are
no geometrical losses due to incomplete coverage of the image plane.
(This is different from some other integral-field instruments:
The PMAS microlens array reimages the exit pupil of the telescope,
and there are therefore no gaps between adjacent fibres.)

The data were reduced using our own IDL-based software package P3d
\citep{becker:02:D}. The reduction consists of 
standard steps such as debiasing and flatfielding using 
twilight exposures, and dedicated routines such as tracing
and extracting the spectra of individual fibres and reassembling
the data in form of a three-dimensional data cube. Wavelength
calibration and rebinning to a constant spectral increment 
is also part of the reduction procedure. 
We extracted the spectra with the iterative fitting method described 
by \citet{wisotzki*:03:IFS}. As the seeing was below 
1\arcsec, the two components of \object{\sds} did hardly overlap and were easy 
to deblend despite the coarse spatial sampling. 
The resulting spectra are shown in Fig.~\ref{fig:specAB}.

\subsection{Comparison of optical spectra}

Already from a superficial examination of Fig.~\ref{fig:specAB} the following
facts can be ascertained: (1) The continuum shapes and slopes of components A 
and B are virtually identical, to the limits of our measurement accuracy.
(2) The emission lines do not have identical shapes. The mismatch is easiest 
seen in the \ion{C}{iv} line at $\lambda\simeq 4230$~\AA, where component A shows 
a much stronger blue line wing than component B; qualitatively the same is 
seen in \ion{He}{ii} at $\lambda\simeq 4480$~\AA,
\ion{C}{iii} at $\lambda\simeq 5220$~\AA, and also in \ion{Si}{iv} at 
$\lambda\simeq 3890$~\AA.
Similar mismatches were already observed in previous spectra of the same target
by \citet{inada*:03} and \citet{richards*:04:ML}. Slit losses in those data,
however, prevented a spectrophotometric comparison of the optical \emph{continuum}.
Richards et al.\ discovered that the strong blue wing excess in the emission lines
of component A over B disappeared between May and Nov 2003, after which the 
A and B spectra were almost (but not quite) identical. Our PMAS data was taken
roughly 4 months after the last epoch covered by Richards et al., and here we
find that a very similar excess feature is again present. We shall discuss below
the implications of this finding in the context of its interpretation as being
due to gravitational microlensing.

It is interesting to compare our new spectra directly with those taken at
earlier epochs. Dr.\ G.T. Richards kindly provided us with their previously
published spectra in digital form. Of special interest is the May 2003 epoch
where the blue wing excess in component A was particularly strong.  Fig.\
\ref{fig:specA_PK} shows the spectral region around the \ion{C}{iv} and
\ion{He}{ii} emission lines for component A, as given by the PMAS and Keck
datasets. In the top panel, the two spectra are directly superimposed, without
any prior rescaling. The two datasets appear to coincide precisely in the
continuum, which however could be accidental,
since neither dataset is of
absolute spectrophotometric quality. In contrast, the \ion{C}{iv} emission
line equivalent width is roughly 40~\% higher in the PMAS data.  The line
\emph{profiles} are more readily compared in the bottom panel, where a
pseudo-continuum (a straight line fitted to narrow spectral windows at
$3920\:\mathrm{\AA}<\lambda<3970\:$\AA\ and
$4625\:\mathrm{\AA}<\lambda<4675\:$\AA) has been subtracted. After scaling the
Keck spectrum by a factor 1.4, the line cores and the red wing of \ion{C}{iv}
match well, but the blue wing is much more enhanced in the Keck spectrum. The
mismatch between PMAS and rescaled Keck data is even more pronounced in
\ion{He}{ii}.

It thus seems that the blue wing excess has disappeared in the second half of
2003, and reappeared, though weaker, in the first quarter of 2004. We are now
interested in the question to what extent this excess flux can be described as
a fixed pattern.  Indeed we find that the residual spectrum $f(A) - S\times
f(B)$, where $S$ is a scaling factor, has a similar shape at both May 2003 and
Apr 2004 epochs.  This is demonstrated in Fig.~\ref{fig:specA_resid} (upper
panel) where the Keck residuals and the somewhat noisier PMAS residuals agree
reasonably well. Another way of pursuing this issue is to ask: Does the blue
wing excess $f(A) - S\times f(B)$ have a similar shape as the epoch difference
in the emission line spectrum of component A alone? And again we find that
this is roughly the case, as shown in the lower panel of
Fig.~\ref{fig:specA_resid}. We summarise that the blue wing excess in
component A appears to be a recurrent feature that changes its strength, but
with a more or less stable spectral shape.

\section{Discussion}
\label{sect:discussion}

\subsection{Microlensing or intrinsic X-ray variability?}

The \xmm\ observations of \object{\sds} reveal  strong discrepancies 
in the optical to X-ray spectral energy distributions of the individual
lens images. Furthermore, the optical spectra of image A
repeatedly showed enhancements of the blue line wings, which so far have never
been observed in any other of the lens images. 
Two different processes can be considered as the reasons for these deviations:
Strong intrinsic flux and spectral variability on timescales comparable to or shorter
than the light path time lags between the images can account for both flux and spectral
deviations.
Microlensing occuring in one or more images can also account in significant flux deviations
in the images and can also change the SED of a lens image, if the emission in different
wavebands arises in regions of different spatial distribution or extent.
This is generally the case for AGN, where the X-ray emission is thought to be emitted 
from a smaller region in the AGN core than the optical emission, which can be partly
attributed to reprocessing in the broad line region.

The fact that both images A and B are so bright in the UV  is almost
certainly due to intrinsic variablity of the QSO, since the two 
images separated by the shortest time delay are equally affected. 
Furthermore it would be very unlikely to find microlensing effects 
in both images.
Also the strong decline of the integral X-ray flux of the object
since the epoch of the ROSAT observation, in conjunction with a significant
hardening of the X-ray spectrum, is evidence for powerful intrinsic variability.

The low A/B ratio in X-rays could in principle also be a consequence of
intrinsic variability coupled with the light path time delay between images A and B,
since AGN variability  is usually strongest and most
rapid in the X-ray regime.
However, images A and B are separated by a time lag of only few weeks 
and at the time of the \xmm observations display a very similar optical to UV SED.
These circumstances make intrinsic variability
a rather unlikely cause for the observed A/B ratio in X-rays.
A microlensing explanation for the low X-ray brightness in component A
requires a demagnification in X-rays of a factor $\sim 3$ relative to
component B, which seems to be rather extreme. 
However, \citet{schechter} calculated microlensing (de-)magnifications 
for the case of a combination of smoothly distributed matter 
(i.e. the cluster dark matter in this case) and the matter in stars.
Their conclusion is that the dilution of microlensing stellar matter
with a smooth component increases the microlensing fluctuations. 
More specifically, their model predicts a high probability of  microlensing 
demagnifications by 1-2 magnitudes, which is exactly the 
demagnification required  here.

The fact that the inverted A/B ratio was still measured by \cxo in January 2005
\citep{ota}, strengthens our case that it is not caused by intrinsic variability,
since the time span between the X-ray observations by far exceeds the possible 
light path time delay between images A and B.
However, the variability of the X-ray A/B ratio, measured at a 2$\sigma$
significance level, may be caused by some intrinsic X-ray variability
or a change in the microlensing configuration.
The persistently harder X-ray spectrum of image D is most easily explained
by intrinsic spectral variability and suggests a time lag of  D with respect to
the other images of more than 9 month.

\subsection{Microlensing of the Broad-Line Region?}

As gravitationally lensed images of the same object the two components A and B should 
have identical spectra, there are only two possibilities to explain the observed 
spectral differences:
On the one hand, the QSO might show rapid intrinsic spectral variability, and the 
light travel time delay causes some such variations to occur in one component 
but not in the other. 
\citet{richards*:04:ML} already made a thoroughly argued case 
against significant and rapid intrinsic variability, on timescales below
the light travel time delay of $\sim 10$--30 days, as the sole origin
of the emission line mismatch. We completely agree and merely note that 
our new observations make this interpretation even less probable, although
it is still a formally possible, however unlikely, option.

Alternatively, component A might be suffering from additional 
microlensing by stars in an intervening galaxy, which could selectively 
amplify certain regions of the source in a way that the QSO spectrum gets modified 
\citep[e.g. ][]{schn+wamb:90,abajas*:02:IGML}. 
This was invoked by \citet{richards*:04:ML} to explain the 
spectral differences between components A and B and the strongly
variable emission line profile of component A. The proposed scenario
is roughly that a caustic (or a critical curve) moved
across only a part of the QSO broad-line region (BLR), amplifiying this 
particular set of clouds so much that the overall line profile changed
temporarily.

A natural prediction of this scenario is that as the caustic pattern continues
to move relative to the source, other parts of the source would get amplified,
causing further spectral variations. Or possibly the caustic could move away
from the compact regions of the source, in which case no more variability
would be expected. The one thing that would most certainly \emph{not} be
expected in this scenario is a reoccurence of the same spectral
feature, without any other major changes -- but
this is exactly what we observed. 
While a cusp caustic could well produce a double-peaked
  high-amplification pattern, there would always be a leading and a trailing
  caustic, both of which would act on different parts of the source.
We also find that the continuum
shapes of components A and B are essentially the same. Thus, the microlensing
amplification would have acted on some inner parts of the BLR (because of the
high velocity differences relative to the line core), but there would be no
chromatic effect on the optical continuum.  We cannot think of a reasonable
source-caustic configuration that would provide such an
amplification pattern.

An alternative explanation of the line variability in \sds was recently  
given by \citet{green}, who proposes variable broad line absorption
by matter surrounding the QSO.
The difference between image A and the other quasar images 
is explained by the small viewing angle differences, resulting in slightly 
different light paths through the intervening matter.
However, a prediction of this model is significant X-ray absorption
with column densities $N_{\rm H} \sim 10^{22} {\rm cm}^{-2}$ in the images
B, C, and D, which is clearly ruled out by the \xmm spectra.

We are thus faced with the situation that no simple, single-cause
explanation will work. We could either try to construct contrived 
models of intrinsic variability where substantial variations always 
happen between the observing gaps. Or we could think of equally contrived
models of caustics and a source shape that produce nearly identical 
differential magnification patterns despite a transverse shift between
caustics and source. We reject both and invoke instead, 
as an act of desperation, a scenario where \emph{both} intrinsic variability 
and microlensing play a role. While our proposed scenario certainly
deserves being called contrived, our only hope is that it may be somewhat 
less contrived than the above mentioned options.

The two main reasons to exclude intrinsic variability as the main driver of 
the differences in the emission lines of \object{\sds} are: (i) The short light travel time delay
in relation to the observed epoch differences; 
(ii) the huge required variability amplitudes in the wings of the broad lines
on these time scales. On the other hand the spectra of component B 
obtained by \citet{richards*:04:ML} show undeniably that at some moderate level, 
the optical spectrum of \object{\sds} does intrinsically change -- unless one invokes
BLR microlensing for that component as well. Furthermore, these spectral changes 
may well occur preferentially in the inner regions of the BLR. Now suppose that
there is indeed a microlens-induced caustic pattern with high magnification power
located such that only a part of the BLR is affected. It would then be that 
particular part of the BLR that gets highly amplified, and the spectrum would
be modified accordingly.
But in contrast to the `pure microlensing' scenario, we do not require this
caustic pattern to move relatively to the source. In fact, we explicitly wish it to
be stationary over the time scales considered. Given the typical effective 
transverse velocities of several 100 km/s, this seems not unrealistic, probably
more realistic than the rapidly moving caustic required in the pure microlensing 
picture. The only cause for spectral variablity would then be intrinsic line
variations, fuelled by the observed X-ray and UV variations.
But these intrinsic line variations would be selectively amplified such that always
the blue wing of the \ion{C}{iv} line, and most of the \ion{He}{ii} line, 
would appear to vary. The Nov/Dec 2003 epoch covered by Richards et al.\
would then have been a period of relative intrinsic quiescence, while May 2003
and Apr 2004 would have seen a slightly more active BLR. 

The possibly best property of this scenario is the fact that it is, in principle,
open to empirical testing. If component A is affected by BLR microlensing and
component B is not, then all variations in A should have been preceeded
(or followed) by much weaker variations in B. The line profile variations need
not be similar if the blue wing excess in A is due to the small size of the
microlensing region, but the \emph{sign} of the variations should essentially 
be the same, and also the relative rates should be in proportion to each other.

\section{Conclusions}

We have presented simultaneous X-ray observations, UV imaging, and  
optical spectroscopy of the first QSO lensed by a cluster of galaxies,
\object{\sds}.
In all observed bands the observations revealed 
significant deviations from 
the simple assumption that all 4 QSO images should have identical 
spectra and spectral energy distributions:
(i) In the optical spectrum of component A  the blue wings of the emission lines
are brighter than in component B. The same pattern had been observed one year earlier
and in the meantime disappeared.
(ii) The UV spectra of components A and B are brighter and harder than 
those of the other components.
(iii) In X-rays the flux ratio A/B is a factor of $\sim 3$ lower 
than in the optical/UV  bands.

We show that neither microlensing nor intrinsic variability can be the single cause
for all observed anomalies.
Instead, we propose a scenario where microlensing is affecting the spectrum of component A.
A part of the broad line region, where the blue wings of the lines are emitted, is magnified 
by microlensing, while the X-ray emitting core of the AGN is demagnified.
However, rather than invoking a moving caustic pattern as cause for the line variability,
we favour a microlensing situation that remains constant over the timespan
of the hitherto performed spectroscopic observations.
In our scenario the line variability is \emph{caused} by flux changes in the UV/X-ray
continuum, but the  variations are much \emph{amplified}  in 
the blue line wings of component A because of microlensing.
The presence of substantial intrinsic variability can hardly be doubted: 
For the broad emission lines this is directly seen in the monitoring 
data by \citet{richards*:04:ML}. For the UV continuum it can be inferred 
from the fact that components A and B have a very similar SED, which in 
turn differs grossly from that of components C and D. Finally, the X-ray
brightness (and spectral shape) has experienced a major change since the 
epoch of the ROSAT observation.

The microlensing scenario in \object{\sds} A is special insofar as this line
of sight is 
far away from the lensing galaxy.  The surface mass density
responsible for the lensing is clearly dominated by smoothly distributed
matter in the cluster, on top of which there comes some action from compact
bodies (in a cluster galaxy or elsewhere). This is thus an extreme case of the
situation envisaged by \citet{schechter}, and we find that the strange
observed relative amplification pattern in this source can be well explained
in their theoretical scenario. 
In particular, the most likely ordering of images in arrival time
of C-B-A-D makes component A a saddle-point image and therefore more
susceptible to microlensing, as was demonstrated by \citet{schechter}.

Due to the relative movement of the intervening galaxy 
the microlensing caustic pattern is bound to change and
the magnifying (and demagnifying) regions of the caustic pattern
will move to  different emission regions of the AGN.
The timescale of this process is governed by the relative transverse velocities
of the AGN, the intervening galaxy, and our observer position with respect 
to each other on the one hand, and the angular scale of the microlensing pattern 
on the other hand. Assuming the intervening galaxy is a member of the lensing
cluster and its velocity is of the order $\sim 1000 {\rm km/s}$, its microlensing
pattern will cross the angle of one stellar mass Einstein-radius in
about 100 years. The magnification patterns calculated by \citet{schechter}
show that the magnifying regions tend to be narrower than the demagnifying regions.

Therefore our proposed scenario predicts for image A that the demagnification of the
X-ray source will remain more or less constant for several years.
Although the magnifying regions of the caustic patterns are smaller, we also expect
the magnification of the blue line wings to change only slowly, since the size of the broad
line region is much larger than the X-ray emitting core.
Hence, our model predicts further occurences of blue line wing excesses
only in image A. However, high SNR spectroscopy should be able to
pick up the underlying line flux variations also in the other components.

Obviously a measurement of the time lags at least between 
images A and B would be extremely helpful to understand the complex
behaviour of the system.
Our finding that the UV spectrum of the QSO is very variable,
suggests that U-band photometry should be a promising method
to measure the time lags.

\begin{acknowledgements}
    GL acknowledges support by the Deutsches Zentrum f\"ur Luft- und Raumfahrt
    (DLR) under contract no.~FKZ 50 OX 0201. LC  acknowledges support by DLR
    grant O5AE2BAA/4 (the ULTROS project). We thank G.T. Richards for making
    his published Keck spectra available to us in digital form. \\
\end{acknowledgements}

\end{document}